# Origin of the springs of Costa Verde beach in Lima, Peru

Rubén Rojas[1], Modesto Montoya[1,2], Enoc Mamani[1], José Maguiña[1], Eduardo Montoya[1], Óscar Baltuano[1], Patricia Bedregal[1], Lucy Coria[2], Alcides Guerra[3], Santiago Justo[3] and Tania Churasacari[3]

[1] Instituto Peruano de Energía Nuclear, Canadá 1470, San Borja, Lima, Perú

[2] Facultad de Ciencias, Universidad Nacional de Ingeniería, Av. Túpac Amaru 210, Lima, Perú

[3] Universidad Ricardo Palma, Av. Benavides 5440, Santiago de Surco, Lima, Perú

**Abstract**

This paper tries to determine the origin of springs on the Costa Verde beach, located in the district of Barranco, Miraflores and Magdalena province of Lima, Peru. These springs emerge near the shoreline, from the lower layers of an 80-meter high cliff. They have survived the process of urbanization of agricultural land, started in the early 70s, which decreased the water table aquifer of Lima, and wiped the water leaks from the cliffs. To identify the source of the springs, isotopic, physical, chemical and bacteriological analysis was carried out for samples from five springs. The isotopic concentrations in waters from Costa Verde's springs are depleted compared to those obtained for Lima aquifer waters, which is recharged by infiltration of the Rimac River. The measured values of those concentrations suggest that water from the Costa Verde's springs should come from a direct recharge in the upper and middle basin, due to infiltration of rainfall or the river at an altitude of about 3600 m. Conductivity and temperature, measured in situ, are similar to those obtained on Lima aquifers. The laboratory analysis showed no significant levels of total or fecal coliform, discarding possible leakage from Lima's sewerage.

*Keywords:* aquifer, water, coast, Lima, Peru

**Introduction**

Between 1989 and 1992, in order to obtain additional hydrogeologic information for better management of water resources for the city of Lima, professionals from Peruvian Nuclear Energy Institute (IPEN) and the Drinking Water and Company of Lima (SEDAPAL) -with the support and advice of the International Atomic Energy Agency (IAEA) in Vienna- studied the origins of the Lima water table [1] and water in Graton tunnel, respectively. The Graton tunnel is located 100 km east from Lima (Peru), at the altitude of 3200 meters [2]. Isotopic analysis was performed in the laboratories of IAEA Vienna, and the chemical analysis was performed in the laboratories of SEDAPAL. In that work it was determined: i) that the Lima aquifer waters come from the infiltration of Rimac and Chillon rivers in the coastal area of the watershed, on identified recharge areas ii) that a portion of the waters from Graton tunnel comes from infiltration of rain in the upper basin of the Mantaro river iii) the altitude gradient of $^2$H and $^{18}$O concentrations in precipitation in the basins of the Rimac and Chillon rivers.

The Graton Tunnel was built by the company Cerro de Pasco Corporation in the late 50s for draining Casapalca mine galleries. This tunnel has a length of 12 km. The mouth, located in the upper basin of the Rimac River, has an altitude of 3200 meters, and delivery to the Rimac river flow ranging between 4 and 6 m$^3$/s. The Rimac and Chillon rivers recharge the aquifer of Lima, which has more than 300 wells drilled by SEDAPAL to provide water for Lima and Callao, cities which together currently have a population of about 10 million.

Having occurred opposite the spa, the process of suppression of springs from the Costa Verde cliff (which is the name given of the beaches of Barranco, Miraflores and Magdalena) was evident. Until the 60s, swimmers still used those springs, that emerge from the 80-meter high cliff, as showers upon leaving the sea. Now there still some springs, near the shoreline. Part of the waters from these springs is used for irrigation of gardens in the area, and the rest is discharged directly into the sea.

In this paper we try to determine the source, hydrodynamics and quality of water from the spring remnants of the Costa Verde Beach. For that, isotopic, physical, chemical and bacteriological analysis is performed.

**Isotopic analysis**

Isotopic analysis of spring waters of the Costa Verde were performed in the laboratories of IPEN with a liquid water isotope analyzer LGR (laser technology), with which we determined the concentrations of stable isotopes $^2$H and $^{18}$O ($\delta^2$H and $\delta^{18}$O, respectively) in waters of those springs. The isotopic concentration results are expressed in δ value in parts per thousand (o/oo) relative to VSMOW (Vienna Standard Mean Ocean Water, standard provided by the IAEA). The results are shown in Table 1.

In Figure 1, $\delta^2$H vs $\delta^{18}$O values, taken from Table 1, together with those obtained in 1990 by Rojas and Ruiz for the Lima aquifer [1], are plotted. It can be seen that the values of $\delta^2$H for springs from the beaches of Costa Verde are between -101.92 o/oo and -108.31 o/oo and the values of $\delta^{18}$O are between -14.41 o/oo and -15.41 o/oo.

These values show an average depletion of $\delta^2$H and $\delta^{18}$O isotopic concentration relative to the values obtained between 1989 and 1990 in the aquifer made in Lima, in the sector recharged by the Rimac river (-102.9 o/oo and -13.97 o/oo, respectively). Those results suggest that a direct recharge comes from a leak of precipitation over an area located at an average altitude of about 3600 meters. This value is deducted from the isotopic gradient, presented on Fig. 2, obtained by Rojas and Ruiz [1] during the years 1989 and 1990. On this figure we can see the relationship between concentration and rainfall height, measured to study the source of the Graton tunnel (personal communication by Ruben Rojas and Juan Carlos Ruiz).

Table 1: Values of concentrations (δ) of $^2H$ and $^{18}O$ and standard deviations (SD) of the water from springs in Costa Verde (composed of beaches of Barranco, Miraflores and Magdalena de Lima, Peru), the basin of Huarangal Chillon river, and the district of Villa Maria del Triunfo, respectively, conducted in November and December 2012.

| Spring | $δ^2H$ Value (0/00) | $δ^2H$ Standard deviation (0/00) | $δ^{18}O$ Value (0/00) | $δ^{18}O$ Standard deviation (0/00) |
|---|---|---|---|---|
| Barranquito | -104.73 | 0.39 | -14.60 | 0.19 |
| Huarangal 1 | -96.36 | 0.59 | -13.01 | 0.03 |
| Huarangal 2 | -95.56 | 0.87 | -13.00 | 0.07 |
| La Estrella | -104.63 | 0.24 | -14.41 | 0.14 |
| Paraiso1 | -96.13 | 0.53 | -13.76 | 0.19 |
| Barranquito | -101.92 | 1.64 | -14.67 | 0.24 |
| La Estrella | -102.77 | 0.36 | -14.78 | 0.15 |
| Los Yuyos | -105.83 | 1.31 | -14.63 | 0.07 |
| Barranquito | -108.31 | 0.09 | -15.01 | 0.05 |
| La Estrella | -105.42 | 0.73 | -15.41 | 0.14 |
| La Estrella | -103.07 | 1.25 | -14.51 | 0.06 |
| Agua Dulce | -102.33 | 1.16 | -14.51 | 0.16 |
| Barranquito | -102.70 | 1.42 | -14.80 | 0.17 |
| Magdalena | -104.86 | 1.09 | -14.64 | 0.12 |

Note: Results in δ(o/oo) referred to Vienna Standard Mean Ocean Water (VSMOW).

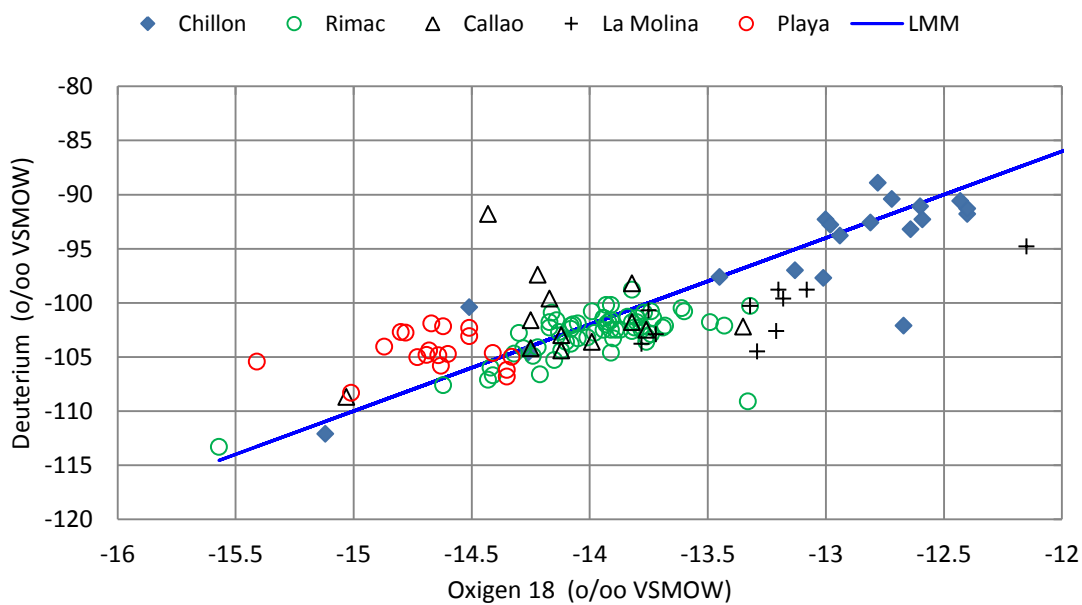

Figure 1: Stable isotope concentrations in aquifer of Lima and waters from springs Lima's Costa Verde

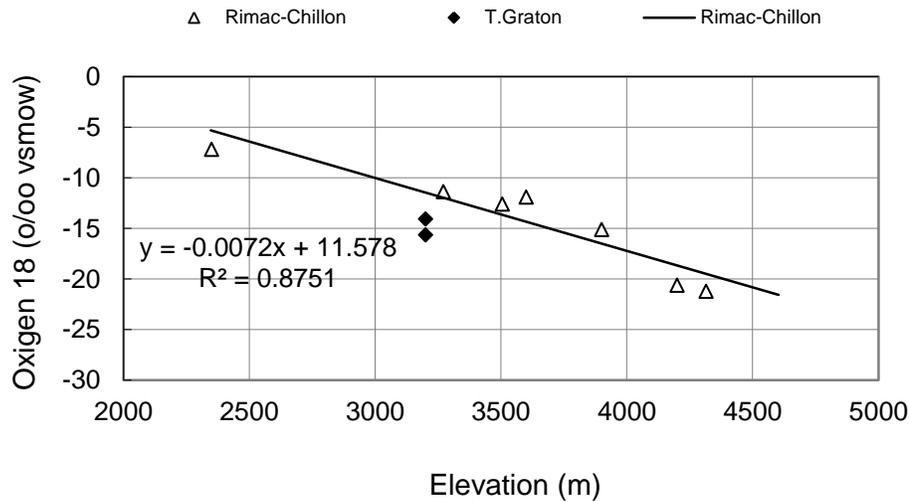

*Figure 2: Isotope concentration gradient for Rimac river and tunnel Graton waters (Personal communication by Ruben Rojas and Juan Carlos Ruiz). Concentrations of $^{18}O$ and the height of the water precipitation are plotted.*

Isotopic concentrations in samples from springs MHUA1 and MHUA2, located near Huarangal Nuclear Center, are compatible with the characteristics of waters of Chillon river basin that they belong. The waters of the springs of Villa Maria del Triunfo (south of Lima) show an origin different from the rest, which requires further studies.

**Physical and Chemical Analysis**

The waters in La Estrella and Barranquito beaches have a temperature of 23°C and 22.5°C, respectively, while sea water has a temperature of 17.7°C. The Barranquito water has a conductivity of 0.77 mS/cm, and that of La Estrella 0.93 mS/cm, while the sea has a very high conductivity that is outside the measuring range of our measure device. See Table 2.

*Table 2: Physical-chemical parameters measured in the field*

| Spring | Conductivity (mS/cm) | Temperature (C) |
|---|---|---|
| Barranquito | 0.77 | 23 |
| Estrella | 0.93 | 22.5 |
| Sea | Out of range of device | 17.7 |

Table 3 presents the values of pH and concentrations (µg/L ± CI) of some elements in samples from the springs of Costa Verde. For analysis, the atomic absorption spectrometry technique has been used. The three samples from Magdalena clearly show arsenic concentrations higher than the other ones. Nevertheless all samples showed lower than the allowable concentrations.

*Table 3. Results of pH and concentrations (μg / L ± CI) measurement of some elements in water from the Costa Verde springs. The values are in the range allowed for human consumption. The source of Magdalena shows an As concentration greater than the other.*

| Spring | pH | As (μg) | Cd (μg) LD=1 | Co (μg) | Cr (μg) LD=.5 | Cu (μg) | Fe (μg) LD=1 | Pb (μg) LD=.5 |
|---|---|---|---|---|---|---|---|---|
| CV Agua Dulce | 7.03 | 1.80 ± 0.10 | ND | < 0.5 | < 0.7 | < 6 | ND | ND |
|  | 6.95 | 2.0 ± 0.3 | ND | <0.5 | ND | <0.6 | ND | ND |
| Los Yuyos | 7.13 | 2.0 ± 0.5 | ND | < 0.5 | < 0.7 | < 6 | ND | ND |
|  | 7.12 | 1.90 ± 0.20 | ND | <0.5 | ND | <6 | ND | ND |
| CV Barranquito | 7.16 | 1.5 ± 0.4 | ND | <0.5 | <0.7 | <6 | ND | ND |
|  | 7.23 | 1.80 ± 0.20 | ND | <0.5 | ND | <6 | ND | ND |
| CV La Estrella | 7.14 | 2.00 ± 0.10 | ND | <0.5 | <0.7 | <6 | ND | ND |
|  | 7.2 | 2.20 ± 0.10 | ND | <0.5 | <0.7 | <6 | ND | ND |
| CV Magdalena | 6.88 | 5.60 ± 0.20 | ND | <0.5 | ND | <6 | ND | ND |
|  | 6.97 | 5.70 ± 0.10 | ND | <0.5 | 0.7 | <6 | ND | ND |
|  | 6.95 | 6.0 ± 0.4 | ND | <0.5 | ND | <6 | ND | ND |

*ND: not detected*

In analytical chemistry laboratory of IPEN the contents of lead, cadmium, mercury and arsenic in the waters of the springs of Villa Maria del Triunfo were measured. We used the technique of "Square Wave Anodic Stripping Voltammetry" (SWASV). The contents of lead, cadmium, mercury and arsenic were, respectively, less than 2 micrograms per liter.

**Bacteriological analysis**

Bacteriological analysis was carried out in the laboratories of the School of Biological Sciences at the University Ricardo Palma. For the determination of total coliform, fermentation 9221B standard technique has been used for water and wastewater analysis [3]. The results are presented in Table 3. Coliforms are 7 per 100 ml for Barranquito spring and 13 per 100 ml for La Estrella Beach.

The number of fecal coliforms was measured with the procedure 9221E for water and wastewater analysis [3]. No fecal coliform was detected in any of the two springs analyzed in season when there is no risk of contamination by bathers.

*Table 4: Results of bacteriological analysis of water from springs Barranquito, La Estrella in Costa Verde beach (made in December 2012) is presented. A well and a spring in district Villa Maria del Triunfo (south of Lima) were also studied.*

| Spring and global position | Day and time | N.M.P Col./100ml Total | Fecals | H |
|---|---|---|---|---|
| Agua Dulce: 12º9´31.71´´S 77º1´35.43´´O. Elevation 2m | 15/02/13 | 4 | 0 | 35 |
| Los Yuyos: 12º9´14.24´´S 77º1´44.42´´O. Elevation 10m. | 15/02/13 | 2 | 0 | 41 |
| Barranquito: 12º8´42.57´´S 77º1´33.37´´O. Elevation 9m. | 18/12/12 | 13 | 0 | 12 |
|  | 15/02/13 | 4 | 0 |  |
| Estrella: 12º8´14.88´´S 77º1´43.77´´O. Elevation 10m | 18/12/12 | 7 | 0 |  |
| Magdalena: 12º6´22.23´´S 77º3´40.63´´O. Elevation 14m. | 24/01/13 | 0 | 0 |  |
| VMT Paraíso1: 12°08'47.52"S 76°56'08.82'O. Elevation 319m | 02/01/13 | >1600 | 319 |  |
| VMT Paraíso2: 12°08'41.82"S 76°55'30.81'O. Elevation. 467m | 02/01/13 | 0 | 500 |  |

*Escherichia coli* contamination has been measured by the standard method 9221 F for the analysis of water and sewer [3]. No evidence of bacterial contamination in the waters of the aquifers in the Costa Verde beach was detected; however water samples of Villa Maria del Triunfo present 1600 total coliform number and 319 fecal coliforms number per 100 ml. The difference in the level of organic pollution of the springs of Costa Verde Beach and Villa Maria del Triunfo, suggests that the latter cross contaminated areas near the surface, which may be due to the high population density with limited services water and drainage that characterize the area.

**Conclusion**

Isotopic analyzes of waters from springs on the Coast of Magdalena and Miraflores districts in Lima Province, Peru, suggest that they have a somewhat different hydrodynamic that the Lima aquifer, and come from leaks of precipitation over areas located to an average altitude of about 3600 msnm. The low or no fecal coliform number and total coliform samples, found the month of December before the summer season, reinforces this hypothesis. The results shown are preliminary and require analysis of more samples at various times of the year, as well as analysis of residence times.

**Acknowledgements**

We are grateful to Wilfredo Cardenas Sulca and Darío Ccaccya Ccaccya, environmental engineering students from the National Technological University south of Lima (UNTECS), for locating and guiding us to the springs Paradise district of Villa Maria del Triunfo, and for his work to promote the preservation of biodiversity in the area. We also are grateful to Julio Kuroiwa Jr., from the National University for Engineers, for his encouraging advices.

**References**

[1] J.C. Ruiz y R. Rojas Molina. Estudio hidrológico del acuífero de Lima (Perú) aplicando técnicas isotópicas, Estudios de Hidrología Isotópica en América Latina 1994. IAEA TECDOC – 635, Organismo Internacional de Energía Atómica, octubre 1995.

[2] A. Plata Bedmar y R. Rojas Molina. Origen de las aguas que drena el túnel Graton. Informe OIEA. INIS-PE-95-02, 1995
[3] Standard Methods for the examination of water and wastewater 21 st Edicion 2005

Email: mmontoya@ipen.gob.pe